\documentstyle[12pt]{article}
\begin{document}
\title{INSTANTANEOUS ACTION AT A DISTANCE IN A HOLISTIC UNIVERSE}
\author{B.G. Sidharth$^*$\\ Centre for Applicable Mathematics \& Computer Sciences\\
B.M. Birla Science Centre, Hyderabad 500 063 (India)}
\date{}
\maketitle
\footnotetext{$^*$E-mail:birlasc@hd1.vsnl.net.in}
\begin{abstract}
The early work of Lorentz, Abraham and others, evolved through the work of
Fokker, Dirac and others to ultimately culminate in the Feynman-Wheeler direct action at a distance theory.
However this theory has encountered certain conceptual difficulties like
non-locality in time, self force of the electron, pre acceleration and the
perfect absorption condition of Feynman and Wheeler, that is the instantaneous
action of the remaining charges in the universe on the charge in question.
More recently, Hoyle and Narlikar have resurrected this theory, but within
the context of a Steady State or Quasi Steady State cosmology. They argue
that the theory infact has a better standing than the generally accepted
quantum theoretic description.\\
In this article we consider a quantum theoretic description and a cosmology
which parallels the Hoyle-Narlikar approach. This leads to a synthesis and
justification of the Dirac and Feynman-Wheeler approaches, clarifying the
conceptual problems in the process. We deduce
a scenario with quantized space-time and a holistic cosmology, consistent
with physical and astrophysical data. The non-locality is now seen to be
meaningful within the minimum space-time intervals, as also the
perfect absorption within the holistic description. Local realism, and the
usual causal field theory are seen to have an underpinning of direct action.
For example this is brought out by the virtual photons which mediate interactions
in Quantum Electro Dynamics, and the emergence of the inverse square law
in the above approach from a background Zero Point Field.
\end{abstract}
\section{Introduction}
When Newton introduced his law of Gravitation in the 17th century, it was
an action at a distance force law. This concept was taken for granted for
about two centuries. In the meantime Roemer by observing the eclipses of the
satellites of Jupiter had come to the startling conclusion that light travels
not at an infinite, but rather at a finite speed.\\
The development of the study of electricity lead to new concepts. Thus Gauss
in the 19th century was already speculating about finite speed transmission
of electrical effects. All this culminated in Maxwell's electromagnetic
theory, in which these effects travelled with the velocity of light.\\
The stage was now set for Einstein to propose his Special Theory of
Relativity. One of the important foundations of Relativity Theory is that of
local realism and related causality, which apparently contradicts the action at a
distance concept.\\
Around the same time Lorentz, Abraham and others were working on the theory
of the electron\cite{r1}, which ultimately ran into several difficulties,
including the problem of infinite self energies arising when the size of
the electron was made to shrink to zero.\\
One would have thought that the action at a distance concept had been buried, but in
1929 Fokker came up with his theory of point particles and direct action.
Though this solved problems like infinite self energy, it could not account
for the phenomenon of radiation.
The theory was further developed by Dirac\cite{r2} in the late thirties who introduced
a term containing a difference of the retarded and advanced fields to
eliminate structure dependent terms. Here also there were problems, like the
run away solutions, the presence of the third derivative of the position, which meant
a non Newtonian Mechanics, self action and so on.\\
In 1945 Feynman and Wheeler\cite{r3} brought the
action at a distance theory to a more acceptable level, in the process
answering the question of lack of radiation in the Fokker theory. However
they had to introduce the sum of the advanced and retarded fields into the
theory and also the perfect absorber condition, that is the action of the
remaining charges of the universe on the charge in question.\\
The topic has
continued to attract attention by several scholars\cite{r4,r5,r6,r7}
like Hoyle and Narlikar, Smirnov-Rueda, Chubyalko and others. On the contrary
this formulation has been refined further in the latest studies.\\
However, concepts like the advanced fields, effectively future to past
propagation and perfect absorption may appear to be unsatisfactory.\\ Moreover
with the development of quantum mechanics, it would also appear that the
entire action at a distance formulation has become irrelevant. But then it
must be borne in mind that quantum theory as it stands may not be the last
word. There are still many unsatisfactory or unexplained features, like the
large number of arbitrary parameters required in the standard model, the
problem of the undiscovered monopole and so on. At the same time the experimentally
well established Quantum teleportation itself undermines the original
concept of local realism and the related concept of causality\cite{r8}.
Today, again the question of superluminal velocities
has been opened up\cite{r9}.\\
The object of the present article is not to go over the formal developments
in action at a distance theory in the recent years. That would be repititious,
and in any case is dealt with in great detail in the references \cite{r1,r5}
and in other articles to appear in this volume. Rather, we will first very briefly
survey the "troublesome" concepts of this theory. Next we will argue that in
the context of the recent work on quantized space time and what may be called
stochastic holism, the entire concept of
instantaneous action at a distance is still meaningful.\\
Indeed Wheeler and others\cite{r10,r11}
recognized that any interaction is non local, or more precisely holistic,
in the sense that an interaction is meaningful, not in isolation, but only
between the interacting systems. This will be seen to have a bearing on the
otherwise strange concepts like that of the perfect absorber condition or
the use of advanced fields.
\section{Instantaneous Action at a Distance}
In this section, keeping in view arguments of the next section, we will
discuss the contributions of Dirac, Feynman and Wheeler. This was built upon
the earlier work of Lorentz, Abraham, Fokker and others. Our starting point
is the so called Lorentz-Dirac equation\cite{r1}:
\begin{equation}
ma^\mu = F^\mu_{in} + F^\mu _{ext} + \Gamma^\mu\label{e1}
\end {equation}
where
$$F^\mu_{in} = \frac{e}{c} F^{\mu v}_{in} v_v$$
and $\Gamma^\mu$ is the Abraham radiation reaction four vector, given by
$$F^\mu_{in} = \frac{e}{c} F^{\mu v}_{in} V_v$$
\begin{equation}
\Gamma^\mu = \frac{2}{3} \frac{e^2}{c^3} (\dot a^\mu - \frac{1}{c^2} a^\lambda
a_\lambda v^\mu)\label{e2}
\end{equation}
Equation (\ref{e1}) is the relativistic generalisation for a point electron of
an earlier equation proposed by Lorentz, while equation (\ref{e2}) is the
relativisitic generalisation of the original radiation reaction term due to
energy loss by radiation. It must be mentioned that the mass $m$ in equation
(\ref{e1}) consists of a neutral mass and the original electromagnetic mass
of Lorentz, which latter does tend to infinity as the electron shrinks to a
point, but, this is absorbed into the neutral mass. Thus we have the forerunner
of renormalisation in quantum theory. It must also be remembered that
equation (\ref{e1}) is valid for a single, that is, isolated electron.\\
We now touch upon three apparently unsatisfactory features of the Lorentz-Dirac
equation (\ref{e1}).\\
Firstly the third derivative of the position coordinate in (\ref{e1})
through $\Gamma^\mu$ gives
a whole family of solutions. Except one, the rest of the solutions are run away -
that is the velocity of the electron increases with time to the velocity of
light, even in the absence of any forces. It is the adhoc assumption of
assymptotically vanishing acceleration that gives a physically meaningful
solution.\\
A second difficulty is that of violation of causality of even the physically
meaningful solutions. Let us see this briefly.\\
For this, we notice that equation (\ref{e1}) can be written in the form\cite{r1},
\begin{equation}
ma^\mu (\tau) = \int^\infty_0 K^\mu (\tau + \alpha \tau_0)e^{-\alpha}
d\alpha\label{e3}
\end{equation}
where
$$K^\mu (\tau) = F^\mu_{in} + F^\mu_{ext} - \frac{1}{c^2}Rv^\mu,$$
\begin{equation}
\tau_0 \equiv \frac{2}{3} \frac{e^2}{mc^3}\label{e4}
\end{equation}
and
$$\alpha = \frac{\tau' - \tau}{\tau_0},$$
It can be seen that equation (\ref{e3}) differs from the usual equation of
Newtonian Mechanics, in that it is non local in time. That is, the
acceleration $a^\mu (\tau)$ depends on the force not only at time $\tau$, but
at subsequent times also. Let us now try to characterise this non locality
in time. We observe that $\tau_0$ given by equation (\ref{e4}) is the
Compton time $\sim 10^{-23}secs.$ So equation (\ref{e3}) can be approximated
by
\begin{equation}
ma^\mu (\tau) = K^\mu (\tau + \xi \tau_0 ) \approx K^\mu (\tau)\label{e5}
\end{equation}
Thus as can be seen from (\ref{e5}), the Lorentz-Dirac equation differs from
the usual local theory by a term of the order of
\begin{equation}
\frac{2}{3} \frac{e^2}{c^3} \dot a^\mu\label{e6}
\end{equation}
the so called Schott term.\\
So, the non locality in time is within intervals $\sim \tau_0$, the Compton time.\\
It must also be reiterated that the Lorentz-Dirac equation must be supplemented
by the asymptotic condition of vanishing acceleration in order to be
meaningful. That is, we have to invoke not just the point electron, but also
distant regions into the future.\\
Finally it must be borne in mind that the four vector $\Gamma^\mu$ given in
(\ref{e2}) can also be written as
\begin{equation}
\Gamma^\mu \equiv \frac{e}{2c} (F^{\mu v}_{ret} - F^{\mu v}_{adv}) v_v\label{e7}
\end{equation}
In (\ref{e7}) we can see the presence of the advanced field implying backward
propagation in time. Infact this term, as is well known directly leads to
the Schott term (\ref{e6}). Let us examine this non local feature.
As is known, considering the time component of the Schott term (\ref{e6}) we
get (cf.\cite{r1})
$$-\frac{dE}{dt} \approx R \approx \frac{2}{3} \frac{e^2c}{r^2}
(\frac{E}{mc^2})^4,$$
whence intergrating over the period of non locality $\sim \tau_0$ we can
immediately deduce that $r$, the dimension of spatial non locality is
given by
$$r \sim c \tau_0,$$
that is of the order of the Compton wavelength. Indeed it is known that this
term represents the effects within the neighbourhood of the charge\cite{r5}.\\
What we have done is that we have quantified the space-time interval of non
locality - it is of the order of the Compton wavelength and time. This will
be relevant to the discussion in the next section.\\
We now come to the Feynman-Wheeler action at a distance theory\cite{r3,r5}.
They showed that the apparent acausality of the theory would disappear if the
interaction of a charge with all other charges in the universe, such that the
remaining charges would absorb all local electromagnetic influences was
considered. The rationale behind this was that in an action at a distance
context, the motion of a charge would instantaneously affect other charges,
whose motion in turn would instantaneously affect the original charge. Thus
\begin{equation}
F^\mu_{ret} = \frac{1}{2} \{ F^\mu_{ret} + F^\mu_{adv} \} + \frac{1}{2}
\{ F^\mu_{ret} - F^\mu_{adv} \}\label{e8}
\end{equation}
The left side of (\ref{e8}) is the usual causal field, while the right side has
two terms. The first of these is the time symmetric field while the second
can easily be identified with the Dirac field above and represents the sum
of the responses of the remaining charges calculated in the vicinity of the
said charge. From this point of view, the self force turns out to be the
combined reaction of the rest of the charges.\\
What is crucial in these calculations is, the property of perfect absorption.
This means that, when an electrical charge is accelerated, all electromagnetic
fields arising therefrom, directly or through interaction of the said charge
with other charges should tend to zero assymptotically, sufficiently rapidly.\\
However within this framework, it is still possible to interchange advanced
and retarted effects to consistently get a radiative reaction that is exactly
opposite to that from the Dirac term. We would then have, mathematically,
a universe with advanced, rather than retarded effects. Wheeler and Feynman argued that it is
the initial conditions that dictate the actual solution.\\
Subsequently it was pointed out that the expansion of the universe, that is
cosmological considerations, rather than adhoc initial conditions suffice to
pick out the actual physical solutions. This work has been carried further, for
example by Hoyle and Narlikar who conclude that the status of the concept of
action at a distance is superior to the status of conventional theory, though
not in the Big Bang model, rather given Steady State or Quasi Steady State
Cosmology\cite{r5,r12}.\\
Inspite of this, the appearance of effects within the Compton space-time intervals,
the obscurity of concepts like perfect absorption and the quantized nature of
radiation, are some of the factors which have lead to the supersession of the
action at a distance approach.
\section{Quantized Space-Time and Stochastic Holism}
We will now put the above considerations in a different, though quantum
mechanical context which makes them meaningful. We first observe that
there have been attempts over the years of describing elementary particles
as mini Black Holes. The logical candidate for an electron would be the
charged and spinning Kerr-Newman Black Hole, which as is well known\cite{r13} reproduces the field
of the electron including the anomalous gyro-magnetic ratio. But the
difficulty has been that for elementary particles, Kerr-Newman Black Holes
would have naked singularities. That is, the radius of the horizon becomes
complex and so, apparently meaningless.\\
More recently it has been shown by the author that this difficulty disappears,
once it is recognised that physics as we know it begins outside the Compton
wavelength and time, $l$ and $\tau$. Indeed, it has been known for a long
time that within these intervals we encounter
the unphysical non-local Zitterbewegung effects, which are symptomatic of precisely
the above fact. Unlike in classical theory, this is acceptable in quantum
theory because of the Heisenberg Uncertainity relation.  As all this has been discussed at length, we merely cite the
various references \cite{r14,r15,r16,r17,r18} and mention the fact that complex coordinates as for the Kerr-Newman electron
or in quantum theory, non Hermitian position operators as for the Dirac
electron disappear once averaging over the minimum space-time units $l$
and $\tau$ is done. To show that indeed the Compton wavelength and time
are minimum cut off or quantized units, we will briefly show that the above
approach leads to a unified description of quarks, electrons and
neutrinos the most fundamental elementary particles.\\
We use the linearized general relativity formula \cite{r13}
\begin{equation}
g_{\mu v} = \eta_{\mu v} + h_{\mu v}, h_{\mu v} = \int \frac{4T_{\mu v}
(t - |\vec x - \vec x'|,\vec x')}{|\vec x - \vec x'|} d^3 x'\label{e9}
\end{equation}
with usual notation and geometrized units $(c = \hbar = 1)$. As is well known
\cite{r13}, the spin angular momentum is given by
\begin{equation}
S_k = \int \epsilon_{klm} x^l T^{mo} d^3 x\label{e10}
\end{equation}
We now consider the microscopic case of elementary particles and Compton scales.
One can easily deduce from (\ref{e10}) that if the boundary of integration
is bounded by the Compton wavelength, then $S_k = \frac{\hbar}{2}$, as for the
electron that is we deduce the non-classical spin half, from classical
considerations.\\
We can also deduce from the above that
\begin{equation}
\frac{ee'}{r} \approx 2Gm\int \eta^{\imath j} \frac{T_{\imath j}}{r}
d^3x'\label{e11}
\end{equation}
where $e' = e$ is the test charge.\\
For the electron one can now deduce the well known fact that\cite{r14}
$$\frac{e^2}{Gm^2} \sim 10^{40}$$
All this was to show briefly how the description of the electron can be
recovered from (\ref{e9}), provided one is outside the Compton
wavelength.\\
As we approach the Compton wavelength however, it was shown that from (\ref{e9}) we get
instead a QCD type potential
\begin{eqnarray}
4 \int \frac{T_{\mu \nu} (t,\vec x')}{|\vec x - \vec x' |} d^3 x' +
(\mbox terms \quad independent \quad of \quad \vec x), \nonumber \\
+ 2 \int \frac{d^2}{dt^2} T_{\mu \nu} (t,\vec x')\cdot |\vec x - \vec x' |
d^3 x' + 0 (| \vec x - \vec x' |^2) \propto - \frac{\propto}{r} + \beta r\label{e12}
\end{eqnarray}
It must also be borne in mind that the usual three dimensionality of space
is typical of spinorial behaviour \cite{r13} and so characteristic of
distances outside the Compton wavelength. At the Compton wavelength scale itself we
encounter low space dimensionality viz., two dimensions and one dimension\cite{r19,r20}.
Using the fact that each of the $T_{\imath j}$ above is given by
$\epsilon /3, \epsilon$ being the energy density, we can see from (\ref{e11})
that at the Compton wavelength, we have the charges $e/3$ and $2e/3$ in
one and two dimensions respectively.\\
To get an idea of the masses encountered at this scale, let us consider the
potential (\ref{e12}) which we multiply by $1/m$ to facilitate comparison
with standard literature, to get
\begin{equation}
\frac{4}{m} \int \frac{T_{\mu v}}{r} d^3 x + 2m \int T_{\mu v} r d^3x \equiv
-\frac{\propto}{r} + \beta r\label{e13}
\end{equation}
For further comparison we consider (\ref{e11}) remembering that at scales
greater than Compton wavelength $m$ is the electron mass, and $e$ is the
electron charge. At the Compton wavelength scale owing to the fractional
charge as seen above, $e^2$ goes over to $\frac{e^2}{10} \sim 10^{-3}$,
so that (\ref{e11}) becomes
$$\frac{10^{-3}}{r} = 2Gm \int \eta^{\mu v} \frac{T_{\mu v}}{r} d^3 x$$
whence,
\begin{equation}
\frac{\propto}{r} \sim \frac{1}{r} \approx 2G.10^3 m \int \eta^{\mu v}
\frac{T_{\mu v}}{r} d^3x\label{e14}
\end{equation}
Comparison of (\ref{e14}) with (\ref{e13}) shows that the now fractionally
charged Kerr-Newman particle has the mass $10^3 m \sim 1 GeV$.\\
Finally it may be remarked that as we encounter predominantly the negative
energy solutions of the Dirac equation at the Compton wavelength, which
have the opposite parity\cite{r21}, so these fractionally charged $1 GeV$
particles display handedness.\\
We have thus deduced all the peculiar properties
of the quarks, at the Compton scale: the identification is complete.\\
We merely mention that in the case of the neutrino, the mass is vanishingly
small so that the Compton wavelength is large, that is by elementary particle
standards, so that again we encounter predominantly the negative energy
components with opposite parity\cite{r22}. Indeed the neutrino is left handed.\\
The above brief review was not only to show the completeness of the concept
of minimum space-time intervals of the order of the Compton wavelength and time in the
context of contemporary quantum physics, but
also to justify it.\\
We now briefly touch upon consequential cosmological deductions\cite{r23}, as this also will
be relevant to the present theme.\\
We use the fact that given $N$ particles, $\sqrt{N}$ particles are
fluctuationally created\cite{r24}. So in view of the above considerations,
$$\frac{dN}{dt} = \frac{\sqrt{N}}{\tau},$$
which on integration leads to,
\begin{equation}
T = \tau \sqrt{N}\label{e15}
\end{equation}
where $T$ is the age of the universe, $N$ is the number of particles and
$\tau$ as before the Compton time of a typical elementary particle namely
the pion. Similarly we can deduce the following relations:
\begin{equation}
R = \sqrt{N}l\label{e16}
\end{equation}
\begin{equation}
\frac{Gm}{lc^2} = \frac{1}{\sqrt{N}}\label{e17}
\end{equation}
\begin{equation}
H = \frac{c}{l} \quad \frac{1}{\sqrt{N}} \approx
\frac{Gm^3_\pi c}{\hbar}\label{e18}
\end{equation}
where $H$ is the Hubble Constant and $m_\pi$ is the pion mass ((\ref{e16})
for example can follow exactly like (\ref{e15}) if space intervals are
used instead of time; it has been deduced alternatively in the literature).
One can verify that all these equations are correct.\\
The model describes an ever expanding universe, as infact latest observations
confirm\cite{r25}. It also deduces from the theory the mass, radius and
the age of the universe, Hubble's law and the Hubble Constant and other
hitherto puzzling relations, like the so called large number
relations. This cosmology is non-singular: there is no Big Bang. Rather, it
shares features of the Prigogine and Steady State models\cite{r15,r23}.\\
We now observe that the equations (\ref{e15}) to (\ref{e18}) are holistic
in that individual or local quantities depend on the large scale parameters of the
universe like $N$ or $R$. Indeed if $\hbar$ and $c$ are considered to be
universal constants then the relation,
$$m_\pi = \frac{\hbar}{lc},$$
in conjunction with (\ref{e16}) shows that the mass itself is holistic
in character. This can also be seen equivalently from (\ref{e18}), which
is a well known empirical but hitherto unexplained and mysterious relation
\cite{r26}. It ties up the mass of a typical elementary particle with the
Hubble Constant! All this goes against Einstein's local realism (and causal
propogation.)\\
The relevance of the above considerations to the action at a distance theory
is the following:\\
We have seen that the theory can be consistently formulated but at the expense
of introducing certain apparently counter intuitive concepts. These are,
the breakdown of causality, self interaction and the instantaneous response
of the rest of the universe in the absorption interpretation of Feynman
and Wheeler.\\
However all these concepts are shown to be meaningful in the above
Quantum Mechanical Kerr-Newman formulation.
As in the case of the Schott-Dirac term, non locality and self interaction take
place within the Compton wavelength and time in the quantized space
time approach, while the holism implicit in relations like equations (\ref{e15}) to
(\ref{e18}) clearly indicate the response of the rest of the universe.\\
Moreover we reiterate that this response is not the causal response of Einstein's local realism.
The fluctuational creation of particles in the interval $\tau$, as expressed
in the equation leading to (\ref{e15}) is instantaneous, as our physical time
intervals begin outside $\tau$.\\
Infact as argued elsewhere Special Relativity itself is valid outside these minimum
intervals. This can be seen easily as follows: Give the minimum space-time
units as above, $\frac{\Delta x}{\Delta t}$ has a maximum value, namely
the velocity of light. So the instantaneous holism referred to above is
acausal only in the approximation in which the minimum space-time intervals
can tend to zero.\\
What all this means is, that the usual causal picture has a holistic, acausal
underpinning, as brought out by, for example, equations (\ref{e15}) to
(\ref{e18}). This point will be touched upon again in Section 4.\\
The other troubling feature of electromagnetism, is the lack of an arrow,
which as pointed out in the earlier section is resolved by introducing the
cosmological expansion of the universe.\\
However in the above model the arrow of time arises quite naturally as
borne out by the equation (\ref{e15}).
\section{Discussion}
1. In the Dirac formulation discussed in Section 2, an isolated electron in
the universe was being considered. There was the concept of non-local self
force, the Schott term (\ref{e6}) and the radiation field, ofcourse with the usual energy
momentum conservation laws. The non-local effects were within Compton time
scales.\\
In the Feynman-Wheeler approach, all the charges in the universe were
considered, the intermediate radiation field was eliminated, and the Schott
term or the Dirac self force turned out to be the action of the rest of the
universe in the vicinity of the charge in question, and not a self force.\\
In our formulation of Section 3 there is non locality not only within the
Compton time but also within the Compton length, both of which represent
minimum or quantized space-time intervals. Further this space-time quantization
is a holistic effect. That is we see a synthesis of the Dirac approach on the
one hand and the Feynman-Wheeler approach on the other.\\
2. The space-time quantization, while allowing instantaneous action within
these intervals, also enables one to overcome the divergences of quantum
field theory\cite{r27}. The parallel with the Hoyle-Narlikar approach
is clear\cite{r5}. There also cosmological boundary conditions provide a natural
cut off that eliminates these divergences.\\
Interestingly also in both these approaches the cosmology is non-singular -
there is no Big Bang, which latter is incompatible with the action at
the distance formulation.\\
Both these cosmologies, that is the fluctuational cosmology discussed in
Section 3 and the Hoyle-Narlikar Steady State or Quasi Steady State cosmology
lead to an ever expanding universe unlike popular Big Bang models. It is
relevant to point out again that latest observations of distant supernovae confirm
this eternal expansion feature.\\
3. However the formulation in Section 3 is stochastic, and one could wonder
how this could lead to results similar to the non-stochastic conventional
approaches. The answer is, via the route of quantum theory as can be seen
from equations (\ref{e15}) and (\ref{e16}), for example.\\
This fact is brought out more transparently by the following argument:\\
The fluctuation in the mass of a typical elementary particle like the pion due to
the fluctuation of the particle number is given by \cite{r24}
$$\frac{G\sqrt{N} m^2}{c^2R}$$
So we have
\begin{equation}
(\Delta mc^2) T = \frac{G\sqrt{N}m^2}{R} T = \frac{G\sqrt{N}m^2}{c}\label{e19}
\end{equation}
as $cT = R$. It can be easily seen that the right side of (\ref{e19}) equals $\hbar$!
That is we have
\begin{equation}
\hbar \approx \frac{G\sqrt{N}m^2}{c},\label{e20}
\end{equation}
The equation (\ref{e20}) expresses the Planck constant in terms of non
quantum mechanical quantities. (Alternatively, (\ref{e20}) is equivalent to the
well known electromagnetism-gravitation ratio given after (\ref{e11})).\\
Further the equation (\ref{e19}) is the
well known quantum mechanical Uncertainity relation,
$$\Delta E \Delta t \approx \hbar.$$
In the light of the above comments we can now see how the inverse square
law of the direct action at a distance theory emerges. In the cosmological
model of Section 3 particles are created fluctuationally from a background
Zero Point Field with the Compton wavelength as a cut off. In this model
the various points are interconnected or form
a network by the background ZPF effects taking place within time intervals
$\hbar/mc^2$ and corresponding to virtual photons of QED. Infact if two
elementary particles, typically electrons, are separated by distance $r$,
remembering that the spectral density of this field is given by\cite{r15},
\cite{r28,r29}
$$\rho (\omega) \alpha \omega^3$$
the two particles are connected by those quanta of
the ZPF whose wave lengths are $\ge r$. So the force of (electromagnetic)
interaction is given by,
$$\mbox{Force} \quad \alpha \int^\infty_{r} \omega^3 dR,$$
where
$$\omega \alpha \frac{1}{R},$$
$R$ being a typical wavelength.\\
Finally,
$$\mbox{Force}\quad \alpha \frac{1}{R^2}$$
Thus in the idealised case of two stationary isolated particles, we have
recovered the Coulomb law. This justifies Feynman's statement that action-
at-a-distance must have a close connection with field theory\cite{r30}. Interestingly
from the above, one could think of action at a distance as due to ZPF quanta
with wavelength equalling the distance between the two particles. (More precisely,
as pointed out in reference \cite{r14}, the Force field is given correctly by
the Kerr-Newman metric).\\
4. Infact, the concept of a field itself implies non-locality or action at
a distance. Thus a field depends on a parametrized time - it denotes all
or much of space at one instant of time\cite{r31}. This aspect has been
discussed elsewhere \cite{r18,r32}. The usual space and time being on the
same footing, as in conventional relativistic theory is true for stationary
states only\cite{r33}. To clarify this aspect further, we would like to point
out that in quantum mechanics, the stationary wave function is real - it
corresponds to the symmetric part only of the causal electromagnetic field
in (\ref{e8}), and there is neither a time arrow nor causality (cf. also
ref.\cite{r34}). When we add the Dirac or antisymmetric part we get
causal fields, the arrow of time and complex wave functions. As seen above,
this term corresponds to holistic effects within the Compton wavelength -
indeed a complex wave function $\psi$ automatically implies bilinear densities
$\psi \psi^*$, that is, we have to average over the small space-time intervals.
It must also be borne in mind that Special Relativity requires unaccelerated
frames. The arrow of time or irreversibility arises
from stochastic processes - for example this is the content of equation
(\ref{e15}). At the microscopic level also, reversibility is approximate. This
is because of the fluctuational nature of $\sqrt{N}$ and the fact that
$\tau$ is non-zero and non-local.\\
5. Finally it is interesting to note that quantized space time leads to
spin and the Dirac equation, which is the origin of the theory of the
electron in quantum mechanics\cite{r35}.
\section{Conclusion}
The original action at a distance theory firstly leads to unphysical effects
within Compton space-time intervals. Thus it has been generally felt that
quantum mechanics supercedes this theory. Secondly, the theory also invokes
the perfect absorption condition which rules out the currently popular Big
Bang cosmology. On the contrary, Hoyle and Narlikar argue that given the
Steady State or Quasi Steady State cosmology, the action at a distance theory
is actually superior to conventional quantum theoretic models.\\
Arguing from a different point of view, we have shown in Section 3 that quantum
theory itself can be formulated in terms of quantized space-time and stochastic
holism, in the light of which, the supposedly unsatisfactory features of the
action at a distance theory become meaningful and the theory itself gets
justified.

\end{document}